\documentclass[%
 reprint,
superscriptaddress,
 amsmath,amssymb,
 aps,
]{revtex4-2}
\usepackage[version=3]{mhchem} 
\usepackage{graphicx}
\usepackage{dcolumn}
\usepackage{bm}
\usepackage{upgreek}
\usepackage{siunitx}

\begin{document}

\preprint{APS/123-QED}

\title{Current-induced switching of thin film $\alpha$-\ce{Fe2O3} devices imaged using a scanning single-spin microscope}

\author{Qiaochu Guo}
\affiliation{School of Applied and Engineering Physics, Cornell University, Ithaca, NY, USA}
\author{Anthony D’Addario}
\affiliation{Department of Physics, Cornell University, Ithaca, NY, USA}
\author{Yang Cheng}
\affiliation{Department of Physics, The Ohio State University, Columbus, OH, USA}
\author{Jeremy Kline}
\affiliation{School of Applied and Engineering Physics, Cornell University, Ithaca, NY, USA}
\author{Isaiah Gray}
\affiliation{School of Applied and Engineering Physics, Cornell University, Ithaca, NY, USA}
\author{Hil Fung Harry Cheung}
\affiliation{Department of Physics, Cornell University, Ithaca, NY, USA}
\author{Fengyuan Yang}
\affiliation{Department of Physics, The Ohio State University, Columbus, OH, USA}
\author{Katja C. Nowack}
\email{kcn34@cornell.edu}
\affiliation{Laboratory of Atomic and Solid State Physics, Cornell University, Ithaca, NY, USA}
\affiliation{Kavli Institute at Cornell for Nanoscale Science, Ithaca, NY}
\author{Gregory D. Fuchs}
\email{gdf9@cornell.edu}
\affiliation{School of Applied and Engineering Physics, Cornell University, Ithaca, NY, USA}
\affiliation{Kavli Institute at Cornell for Nanoscale Science, Ithaca, NY}

\date{\today}

\begin{abstract}
Electrical switching of N\'eel order in an antiferromagnetic insulator is desirable as a basis for memory applications. Unlike electrically-driven switching of ferromagnetic order via spin-orbit torques, electrical switching of antiferromagnetic order remains poorly understood. Here we investigate the low-field magnetic properties of 30 nm thick, c-axis oriented $\alpha$-\ce{Fe2O3} Hall devices using a diamond nitrogen-vacancy (NV) center scanning microscope. Using the canted moment of $\alpha$-\ce{Fe2O3} as a magnetic handle on its N\'eel vector, we apply a saturating in-plane magnetic field to create a known initial state before letting the state relax in low field for magnetic imaging. We repeat this procedure for different in-plane orientations of the initialization field. We find that the magnetic field images are characterized by stronger magnetic textures for fields along $[\bar{1}\bar{1}20]$ and $[11\bar{2}0]$, suggesting that despite the expected 3-fold magneto-crystalline anisotropy, our $\alpha$-\ce{Fe2O3} thin films have an overall in-plane uniaxial anisotropy. We also study current-induced switching of the magnetic order in $\alpha$-\ce{Fe2O3}. We find that the fraction of the device that switches depends on the current pulse duration, amplitude and direction relative to the initialization field. Specifically, we find that switching is most efficient when current is applied along the direction of the initialization field.

\end{abstract}

\maketitle

\section{\label{sec:intro}Introduction}
Antiferromagnetic (AF) materials are interesting for future memory and logic applications due to their sub-picosecond spin dynamics~\cite{Jungwirth2016, Jungfleisch2018, Baltz2018} and their potential for high-density information storage~\cite{Loth2012}. Because AF materials have a very small or zero net magnetic moment, direct magnetic manipulation of N\'eel order is not practical for applications. An attractive approach is to switch N\'eel order \emph{electrically}, such as using spin-orbit torques (SOTs) generated at the interface between the AF and a heavy metal layer. SOT-induced N\'eel order switching has been reported in the AF metals CuMnAs~\cite{Wadley2016, Grzybowski2017} and \ce{Mn2Au}~\cite{Z2014, Meinert2018}. Several studies have also shown current-induced N\'eel order switching in AF insulators including CoO~\cite{Baldrati2020}, NiO~\cite{Meer2021, Baldrati2019, Moriyama2018, Chen2018} and $\alpha$-\ce{Fe2O3}~\cite{Cheng2020, Zhang2019, Zhang2022, Cogulu2021} by detecting changes in the Hall resistivity. However, purely electrical measurements are difficult to interpret. For example, electromigration can influence the measured signal in a way that mimics a switching-like response~\cite{Chiang2019, Churikova2020}. Magnetic imaging techniques have been used to reveal the local magnetic order switching in AF insulators, including birefringence imaging of NiO~\cite{Meer2021}, X-ray magnetic linear dichroism (XMLD), photoemission electron microscopy (PEEM) of NiO ~\cite{Baldrati2019, Moriyama2018} and $\alpha$-\ce{Fe2O3}~\cite{Cogulu2021}, and spin Seebeck microscopy of NiO~\cite{Gray2019}. Imaging studies of N\'eel order switching in both NiO and $\alpha$-\ce{Fe2O3} have revealed current-induced switching in SOT-free regions of the sample~\cite{Meer2021, Cogulu2021}, suggesting SOT is either not or not solely responsible for switching. At the current densities necessary to produce strong SOTs, the substantial Joule heating can induce thermal expansion and thus strain in the AF layer. Strain can also switch the N\'eel order without SOT via magnetoelastic coupling~\cite{Meer2021}. 

To investigate the magnetism and associated phenomenology of current-induced N\'eel order switching in an AF insulator, especially considering the influence of the potentially complicated, nanoscale domain order present in these materials, we use a diamond nitrogen-vacancy (NV) center scanning microscope to image the magnetic order in the canted AF insulator $\alpha$-\ce{Fe2O3}. We perform two experiments to study the influence of magnetic field and electric current on the magnetic order of $\alpha$-\ce{Fe2O3}. First, we study how the orientation of an initializing in-plane magnetic field influences the resulting magnetic state in a low field. We find that the final magnetic state of the sample is influenced by the initialization field direction. In particular, our results are consistent with an in-plane uniaxial anisotropy with N\'eel vector easy axis $[\bar{1}100]$ rather than the expected threefold magneto-crystalline anisotropy. Second, we apply current pulses through the Pt layer to investigate current-induced switching. We find that the magnetic order of $\alpha$-\ce{Fe2O3} can be switched with current pulses and that the switching efficiency is determined by the current pulse duration, amplitude, and direction relative to the initialization field. We discuss this finding in the context of possible switching mechanisms. 

\section{\label{sec:methods}Methods}
\begin{figure}[!h]
\includegraphics[scale=1]{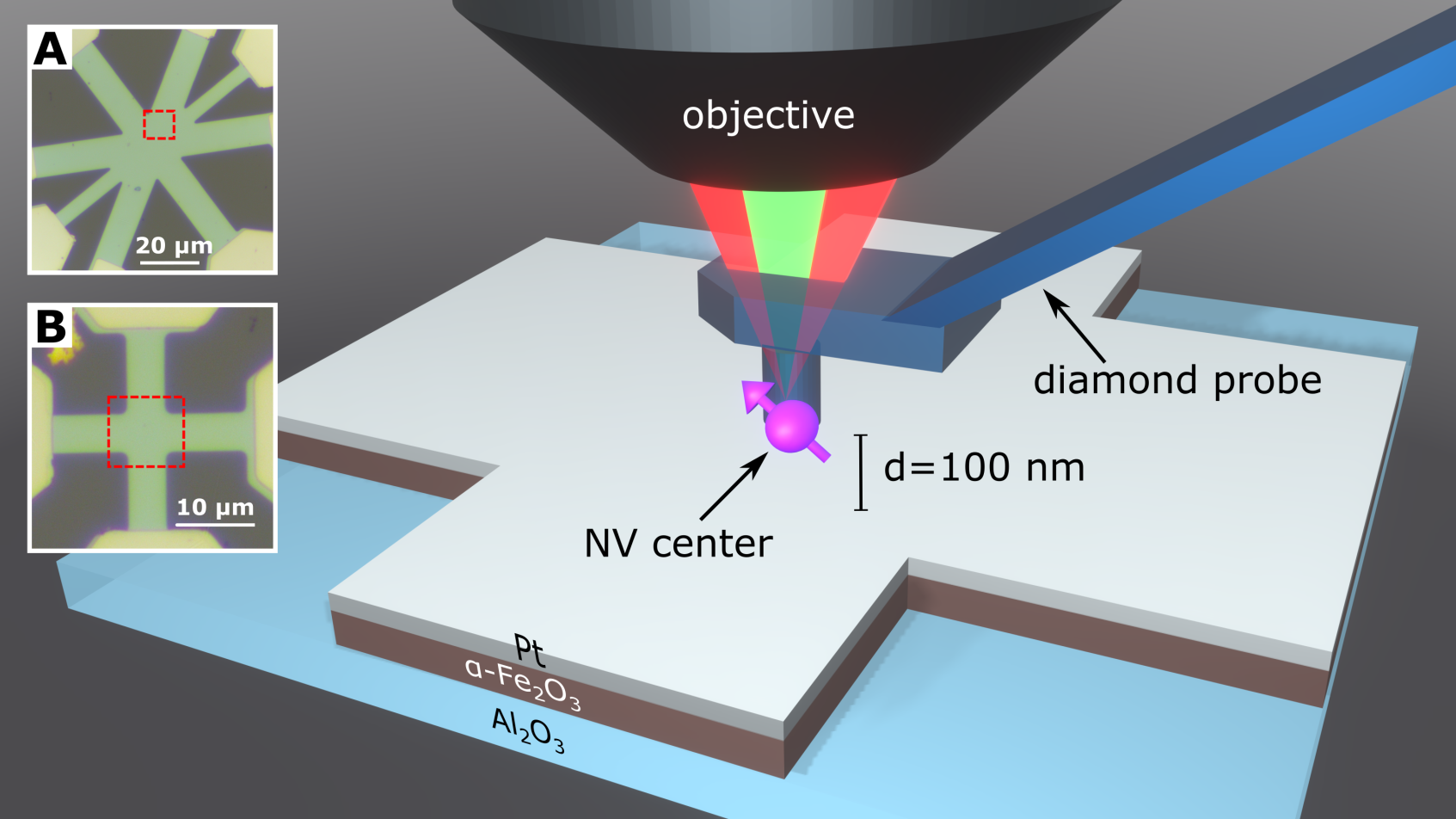}
\caption{\label{fig:setup} Schematic of the scanning NV center setup. We use a commercial diamond probe (QZabre LLC) with a single NV center implanted approximately 10~nm below the tip surface. Scans are obtained with a probe-to-sample separation of 100 nm. A microscope objective is used to focus the green excitation beam and collect the red photoluminescence from the NV center. Insets show optical images of device A (top) and B (bottom). The Pt capped $\alpha$-\ce{Fe2O3} appears bright, and the bare \ce{Al2O3} substrate dark. The red boxes indicate the \SI{10}{\um} $\times$ \SI{10}{\um} scan area.}
\end{figure}

We use a home-built scanning NV microscope to image the stray magnetic field above $\alpha$-\ce{Fe2O3} devices (Fig.~\ref{fig:setup}). Diamond NV centers are sensitive nanoscale magnetometers~\cite{Hong2013,Schirhagl2014,Rondin2014,Hopper2018} that have been used to image  magnetic materials including AFs~\cite{Maletinsky2012,Rondin2012,Casola2018,Gross2017, Chauleau2019, Haykal2020,Wornle2021}, 2D materials~\cite{Thiel2019}, and materials that host skyrmions~\cite{Dovzhenko2018,Gross2018}. Antiferromagnetic spin wave dynamics have also been studied using NV centers~\cite{Wang2022}. Our devices are fabricated from a 30~nm thick epitaxial $\alpha$-\ce{Fe2O3} film grown on a \ce{Al2O3}(001) substrate by off-axis sputtering. 6~nm of Pt is deposited on top of the $\alpha$-\ce{Fe2O3} layer \emph{in situ} at room temperature. $\alpha$-\ce{Fe2O3} has canted spin-order that leads to a weak saturation magnetization of approximately 2~emu/\SI{}{\cm^3}. This moment provides a fringe field that is easily detectable using an NV center as a magnetometer. The entire film stack is patterned into Hall devices to enable electrical measurements. Figure~\ref{fig:setup} shows optical images of two devices that we study; data from an additional device is included in the SI. Device A has six \SI{10}{\um}-wide leads and two \SI{5}{\um}-wide leads, while device B has four \SI{5}{\um}-wide leads. 

We measure the local magnetic field at the NV center, which is integrated into a scanning probe (Fig.~\ref{fig:setup}) using optically detected magnetic resonance (ODMR). The NV center spin resonance frequency is sensitive to the magnetic field component parallel to the NV axis, which for our probes is oriented at a 54$^{\circ}$ angle with respect to the sample-plane normal and with an in-plane projection along $[\bar{1}100]$. All measurements are made using a $\sim$20~G bias magnetic field oriented along this direction, with the NV center scanning 110-nm above the sample surface. The microwave excitation field is applied either by driving a microwave current directly through the Pt layer or by a printed circuit board (PCB) resonator below the sample~\cite{Sasaki2016}. We use two methods to obtain  magnetic images (see SI section II for details): the dual-iso-B method~\cite{Rondin2012} and the resonance frequency tracking method ~\cite{Schoenfeld2011, Welter2022}. In dual-iso-B~\cite{Rondin2012}, we excite the NV center at two fixed microwave frequencies and measure the difference between their respective photoluminescence (PL) values. From a reference ODMR spectrum, we can calculate the NV center resonance frequency shift, and thus the local magnetic field change. This method, however, is limited to the $\sim$1.5~G ODMR linewidth. To avoid saturation, we also use resonance frequency tracking~\cite{Schoenfeld2011, Welter2022}, which adjusts the microwave frequencies at every pixel to track the ODMR peak. All measurements are performed under ambient conditions. 

\section{\label{sec:RnD}Results and Discussion}
\subsection{\label{sec:Binit}Magnetic field initialization}
\begin{figure}[!ht]
\includegraphics[scale=1]{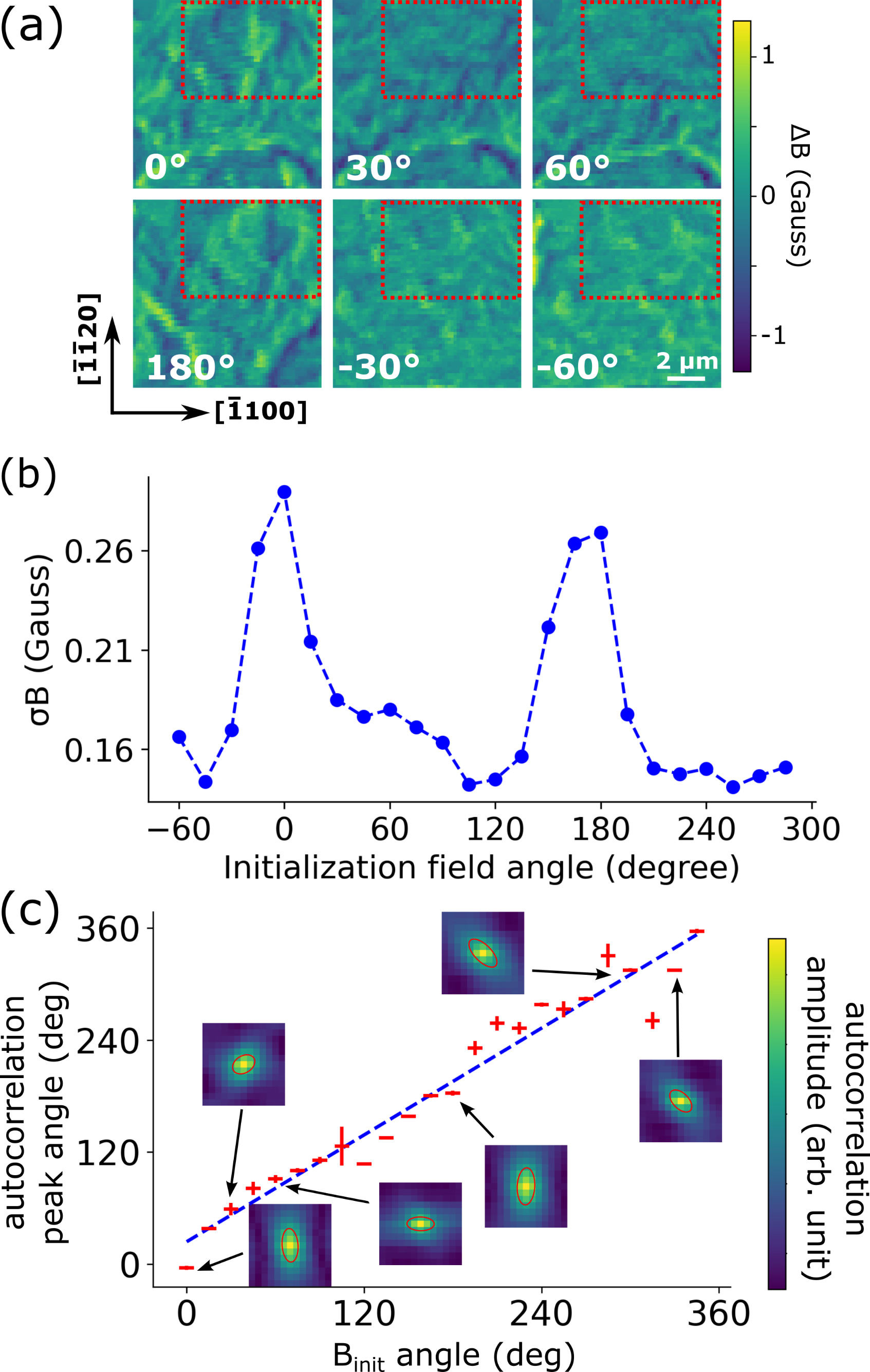}
\caption{\label{fig:Binit} (a) Examples of magnetic images from device A after it was initialized with a 1~T in-plane magnetic field. The in-plane angles of the initialization field with respect to the vertical axis of the image are given in the bottom left corners. Autocorrelation images are calculated from the red-boxed regions. (b) Magnetic field standard deviations of the post-initialization images  as a function of the initialization field angle. Peaks are visible for 0$^\circ$ and 180$^\circ$ corresponding to $[\bar{1}\bar{1}20]$ and $[11\bar{2}0]$. (c) The fitted autocorrelation peak angle as a function of the initialization field angle. The linear fit has a slope of 0.95 and an offset of 23.8$^\circ$, with an R$^2$ value of 0.96. 
}
\end{figure}

First, we study the magnetic field initialization of the sample. Since current-induced magnetic switching measurements are performed at low field, it is important to investigate the sample's low-field magnetic state, how it depends on history, and if it is influenced by magnetocrystalline anisotropy. We apply a 1~T in-plane magnetic field, then ramp the field to zero and move the sample into the scanning NV microscope to image the magnetic field above the sample using resonance frequency tracking. We repeat this measurement, rotating the initialization field in-plane by 15$^\circ$ over a total of 360$^\circ$ to study the influence of the initialization field direction on the sample's magnetic state. Figure~\ref{fig:Binit}(a) shows examples of the corresponding magnetic images taken on device A. The magnetic images reveal strong features that suggest the presence of a magnetic domain wall, but also more subtle textures reminiscent of magnetic ripples. The images appear qualitatively similar, including the re-nucleation of a domain wall in nearly the same position, suggesting there is strong pinning that influences the relaxation of the magnetic state from the high-field saturated state to a low-field multi-domain state.  

The real-space images reveal that, compared to other angles, the magnetic textures are stronger when the initialization field is applied along 0$^\circ$ and 180$^\circ$, i.e. parallel to the $[\bar{1}\bar{1}20]$ direction. To quantify this observation, we calculate the standard deviation of the magnetic field pixels inside the red-boxed regions to avoid the magnetic domain wall feature, which otherwise dominates the result. The result shows pronounced peaks at 0$^\circ$ and 180$^\circ$ (Fig. \ref{fig:Binit}(b)). The 180$^\circ$ periodicity suggests uniaxial magnetocrystalline anisotropy. The stronger magnetic texture observed when the initialization field is parallel to $[\bar{1}\bar{1}20]$ further suggests that the N\'eel vector easy axis is along $[\bar{1}100]$.  This is in contrast to previous suggestions of threefold anisotropy in $\alpha$-\ce{Fe2O3} \cite{Cheng2020}, and is consistent with a previous report of uniaxial anisotropy\cite{Zhang2019}, except with the N\'eel vector easy axis along $[\bar{1}2\bar{1}0]$, 60 degrees away from our observation. Such sample-to-sample variation may be explained by built-in stresses in thin-film samples.

To understand the initialized magnetic states, it would be ideal to reconstruct the sample magnetization directly from the magnetic images. However, reconstruction of the magnetization from a magnetic image is an underconstrained problem~\cite{Casola2018}. Without prior knowledge about the sample's magnetization, for example, whether it is oriented in- or out-of-plane, many solutions may exist that correspond to the same experimental magnetic image. The orientation of the magnetization in $\alpha$-\ce{Fe2O3} thin films is still unclear. Previous studies~\cite{Cheng2020, Zhang2019} assume that the N\'eel vector for c-axis $\alpha$-\ce{Fe2O3} thin films is purely in the sample plane. However, recent x-ray microscopy has revealed that thin films similar to ours also have an out-of-plane magnetic component~\cite{Cogulu2021}. Therefore, direct reconstruction will not be single-valued and may not be reliable.

Instead, we further analyze the magnetic textures by calculating the 2-dimensional autocorrelation~\cite{Reith2017,NavaAntonio2021}
\[R(\delta x, \delta y)=\sum_{x,y}I(x, y)\cdot I(x+\delta x, y+\delta y),\]
where $R$ is the autocorrelation value; $\delta x$ and $\delta y$ are the displacements from the corresponding x and y, and $I(x, y)$ is the pixel intensity at $(x, y)$. The autocorrelation measures the average correlation between one pixel and its surrounding pixels at varied distance. The anisotropy of the center peak of the autocorrelation (see insets in Fig.~\ref{fig:Binit}(c)) indicates the preferential orientation of features in the original image. We compute the autocorrelation in the regions marked with red boxes and perform ellipse fitting on contours of the autocorrelation peaks to calculate their rotation angle relative to the vertical direction. Surprisingly, the angle has a linear relationship with the initialization field angle (Fig.~\ref{fig:Binit}(c)), suggesting that the sample retains a memory of the direction of the initialization field.  

\subsection{\label{sec:current}Current-induced magnetic order switching}
Next, we study how current pulse duration, amplitude, and direction influence current-induced magnetic switching. We first pass a single DC current pulse through the Pt layer of device B using two opposing contacts as source and drain. We then connect the same contacts to a signal generator to supply the microwave magnetic field needed for magnetic imaging. We acquire a magnetic image using the dual-iso-B method after each current pulse and repeat this process for different current pulse durations and amplitudes. This sequence is performed with three combinations of current and initialization field directions. We subtract the initial magnetic image from the images taken after passing a current pulse to obtain difference images showing local changes of the magnetic field that indicate N\'eel order switching due to the applied current pulses.

\begin{figure*}[!ht]
\includegraphics[scale=1]{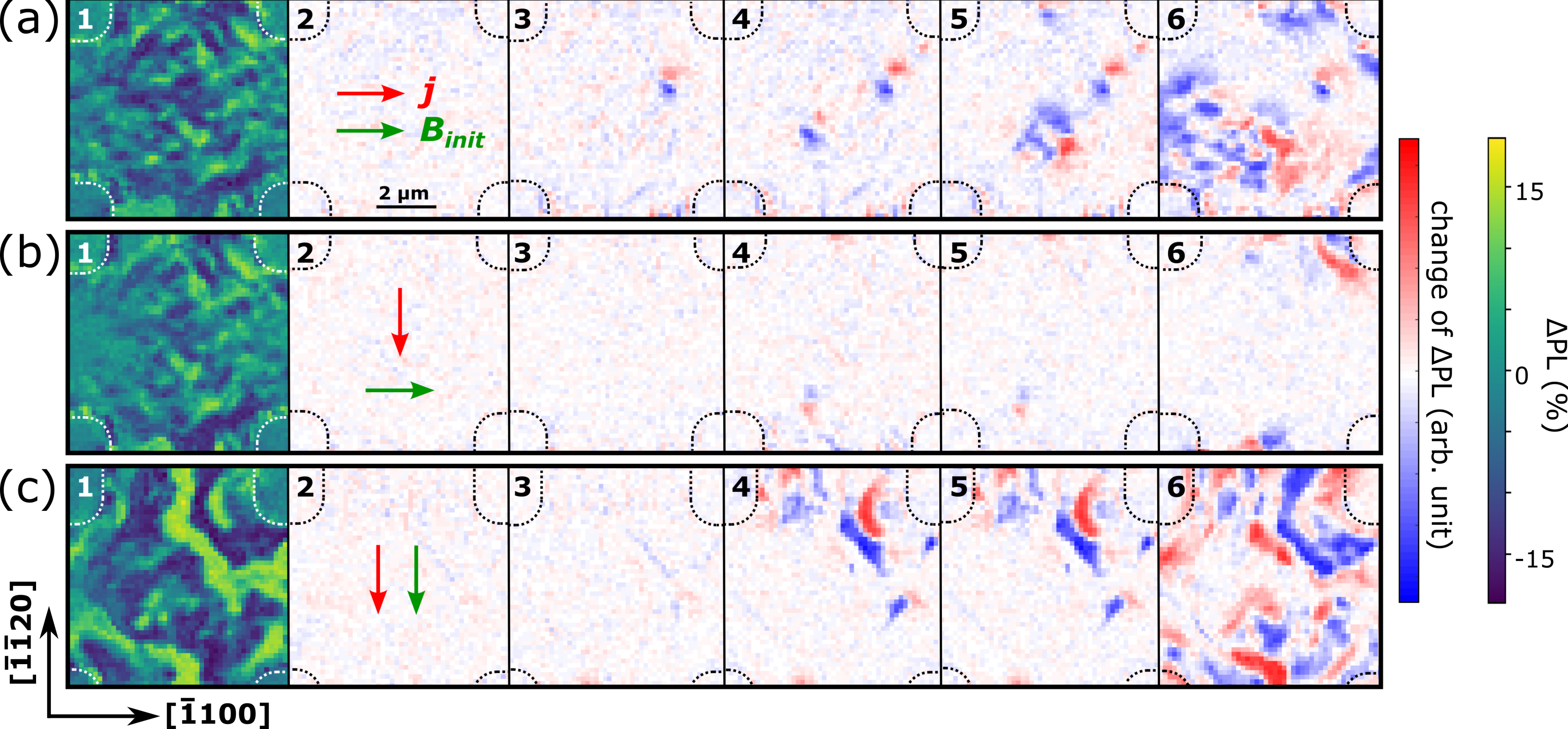}
\caption{\label{fig:diff} Magnetic field images acquired using the dual-iso-B method. (1) Image before applying any current pulse. (2-6) Difference images obtained by subtracting the scan in (1) from the scan taken after passing \SI{100}{\us} current pulses with an amplitude of (2) 3~mA, (3) 7~mA, (4) 13~mA, (5) 16~mA, and (6) 21~mA. The green arrows indicate the direction of the initialization field and the red arrows indicate the current pulse direction. }
\end{figure*}

First, we study the effect of current pulse amplitude. We use a pulse duration of \SI{100}{\us} and vary the amplitude from 3~mA ($j = 1 \times 10^{11}$ A/m$^2$) to 21~mA ($j = 7 \times 10^{11}$ A/m$^2$) for two initialization field directions and two current directions. Figure~\ref{fig:diff} shows the resulting magnetic field difference images. For all orientations, a larger fraction of the sample is switched as the current amplitude increases. Comparing the three different combinations of current pulse and initialization field directions (arrows in Fig.~\ref{fig:diff}(2)), we see that a larger fraction of the sample switches when the current and the initialization field are in the same direction (Fig.~\ref{fig:diff}(a) and (c)) than when the two are perpendicular to each other (Fig.~\ref{fig:diff}(b)). 

Our field initialization experiment has shown that the initialization field direction influences the resulting low field magnetic texture. A reasonable speculation is that the canted moment is oriented primarily in the initialization field direction. If so, our observation that a larger fraction of the sample switches when the current and initialization field are parallel indicates that the current pulses tend to rotate the canted moment out of the current direction, i.e. align the N\'eel order with the current direction. This is consistent with results from previous Hall resistivity measurements~\cite{Cheng2020,Zhang2019}. Similar switching behavior is also observed by Meer et al~\cite{Meer2021} in Pt/NiO as a result of a thermal-magnetoelastic effect. This study suggests that the direction of magnetoelastic-effect-induced switching depends on the geometry of the current path, and in a 4-leg device similar to ours, current pulses passing through two opposing contacts rotate the N\'eel vector to the current direction~\cite{Meer2021}. 

Finally, we study the influence of the current pulse duration on switching. We apply 10~mA current pulses and vary the pulse duration from \SI{1}{\us} to \SI{500}{\us}. We observe that a \SI{1}{\us} current pulse does not induce switching. To confirm this observation, we increase the current amplitude to 25~mA ($j = 8.3 \times 10^{11}$ A/m$^2$), while maintaining the pulse duration as \SI{1}{\us}, and still do not find magnetic order switching (Fig.~S7). The absence of current-induced switching with a \SI{1}{\us} pulse suggests that switching may require thermal activation, and that it may be a slow process. A 10~mA, \SI{10}{\us} pulse induces a small amount of switching, but extending the pulse duration further does not lead to additional significant changes. We also observe switching outside the current path after passing strong current pulses (Fig.~\ref{fig:diff}(a-6) and (c-6), Fig.~S8), where there is no SOT. These results further support that thermal-magnetoelastic effects are sufficient to produce magnetic switching, and that SOT may not be the most important mechanism for current-induced switching in our samples. 

\section{\label{sec:conclusions}Conclusions}
To conclude, we image the stray magnetic field above a thin-film $\alpha$-\ce{Fe2O3} device using scanning NV microscopy. The NV images suggest that the orientation of an in-plane 1~T magnetic field influences the sample's magnetic state even after relaxation in a low field. When the initialization field is along the $[\bar{1}\bar{1}20]$ or $[11\bar{2}0]$ direction, the resulting state has more pronounced magnetic texture reflected in a larger standard deviation in the magnetic field image. This suggests an in-plane uniaxial anisotropy in our $\alpha$-\ce{Fe2O3} sample with N\'eel vector easy axis along $[\bar{1}100]$. N\'eel order can be switched by applying a current pulse through the thin Pt layer deposited on the $\alpha$-\ce{Fe2O3} layer. We observe no current-induced magnetic switching with current pulses shorter than \SI{1}{\us}, and once the sample is switched, extending the pulse duration does not induce additional switching. We observed current-induced switching outside the current path. These results indicate that SOTs may not be necessary to induce magnetic switching in $\alpha$-\ce{Fe2O3}. We also find that increasing the current amplitude leads to a larger switched area, with the most efficient switching when the current pulse is in the same direction as the initialization field direction.

\begin{acknowledgements}
This work is primarily supported by the National Science Foundation (DMR-2004466).  Quantitative peak tracking was developed with support by the U.S. Department of Energy, Office of Science, National Quantum Information Science Research Centers (1F-60510). The PCB-based microwave resonator was developed with support from the U.S. Department of Energy, Office of Science, Basic Energy Sciences (DE-SC0019250). The development of the scanning NV microscope set-up was supported by the Cornell Center for Materials Research (CCMR) with funding from the NSF MRSEC program (DMR-1719875), including capital equipment support by CCMR and the Kavli Institute at Cornell. Sample growth is supported by the Department of Energy (DOE), Office of Science, Basic Energy Sciences (DE-SC0001304).

\end{acknowledgements}

\bibliography{main_ref}
\bibliographystyle{apsrev4-1}
\end{document}


\preprint{APS/123-QED}

\newcommand*\mycommand[1]{\texttt{\emph{#1}}}
\renewcommand{\thefigure}{S\arabic{figure}}

\author{Qiaochu Guo}
\affiliation{School of Applied and Engineering Physics, Cornell University, Ithaca, NY, USA}
\author{Anthony D’Addario}
\affiliation{Department of Physics, Cornell University, Ithaca, NY, USA}
\author{Yang Cheng}
\affiliation{Department of Physics, The Ohio State University, Columbus, OH, USA}
\author{Jeremy Kline}
\affiliation{School of Applied and Engineering Physics, Cornell University, Ithaca, NY, USA}
\author{Isaiah Gray}
\affiliation{School of Applied and Engineering Physics, Cornell University, Ithaca, NY, USA}
\author{Hil Fung Harry Cheung}
\affiliation{Department of Physics, Cornell University, Ithaca, NY, USA}
\author{Fengyuan Yang}
\affiliation{Department of Physics, The Ohio State University, Columbus, OH, USA}
\author{Katja C. Nowack}
\email{kcn34@cornell.edu}
\affiliation{Department of Physics, Cornell University, Ithaca, NY, USA}
\author{Gregory D. Fuchs}
\email{gdf9@cornell.edu}
\affiliation{School of Applied and Engineering Physics, Cornell University, Ithaca, NY, USA}

\title  {Supplemental material: Current-induced switching of thin film $\alpha$-\ce{Fe2O3} devices imaged using a scanning single-spin microscope}

\maketitle
\section{Experimental Methods}

\subsection{Brief summary of relevant NV physics}
\begin{figure}[!htbp]
\includegraphics[scale=0.5]{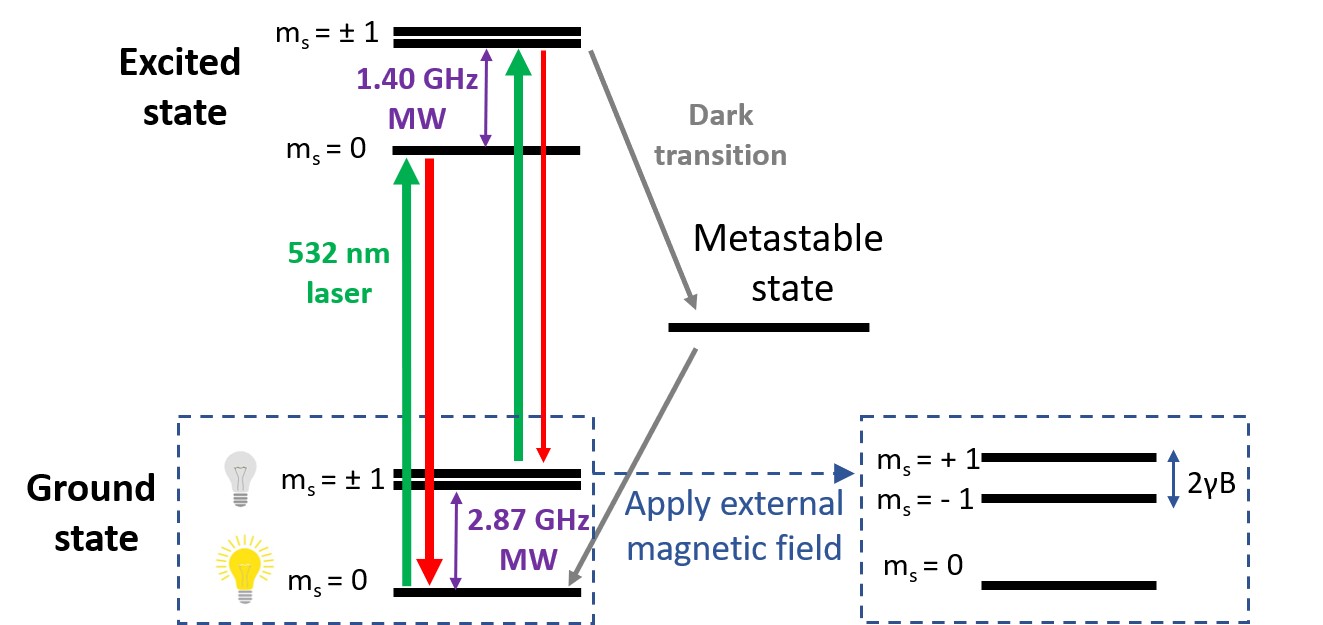}
\caption{\label{fig:NVenergy} NV energy level diagram. The green arrows represent laser excitation, and the red arrows represent the NV PL. From the $m_s=\pm1$ states, the NV center has a chance to go through a dark transition and end up emitting less fluorescence.}
\end{figure}

Nitrogen-vacancy (NV) centers in diamond are spin-1 point defects that can be excited with a 532 nm optical excitation and read out through photoluminescence (PL) that depends on the spin state. The PL intensity measured from an NV center depends on its spin state: NV centers in the $m_s=\pm1$ states have a finite probability of going through a metastable state and returning to the $m_s=0$ state. This transition does not emit visible light (Fig.~\ref{fig:NVenergy}). As a result, we can read out the NV center's spin state by measuring its PL. With an external magnetic field, the degenerate $m_s=\pm 1$ states are split due to the Zeeman effect. We take optically-detected magnetic resonance (ODMR) spectra of an NV center by measuring its PL while sweeping the microwave frequency. The magnetic field along the NV axis can be directly calculated from the frequency difference between the two dips on the ODMR spectrum\cite{Hong2013,Schirhagl2014,Rondin2014,Hopper2018}. 

\subsection{Scan NV setup}
Our homebuilt scanning NV microscope is based on a tuning fork atomic force microscope (AFM) from Mad City Labs including a customized 3-axis piezostage. We purchase NV-diamond scanning probes from QZabre LLC. The probe tip is fabricated from [100]-cut diamond, and a single NV center is implanted approximately 10 nm below the probe tip surface, with the N-V axis oriented along the [111] direction. Figure~\ref{fig:singleESR} shows an ODMR spectrum of a probe with a hyperfine splitting of ~3.1 MHz caused  by $^{15}$N impurities around the NV center. A 532 nm green laser is focused on the NV center through a 100x, 0.9 NA microscope objective, and the PL from the NV center is collected through the same optical path. We use an avalanche photodiode (APD) to count the number of red photons emitted from the NV center in a chosen time window and convert this number to count rate in kCounts/s. A permanent magnet is placed nearby the scanning stage, providing a bias field of 10 to 20 Gauss that is roughly aligned to the NV-axis. We use two methods to apply the microwave (MW) excitation: 1) directly passing a high frequency current through the Pt layer on the sample, or 2) exciting a printed circuit board (PCB) resonator based on a design by Sasaki et al.\cite{Sasaki2016} on which samples can be directly mounted. When using the first method, two legs of a Hall-cross are connected to the signal generator to drive the high frequency current through the sample. Due to the spatial structure of the resulting MW magnetic field, the MW power exciting the NV center is spatially non-uniform resulting in lower contrast in regions that are outside the MW path. The PCB resonator produces a more uniform MW magnetic field. The sample is directly mounted on the PCB, and NV images are taken within the region above the resonator's central circle. 

\subsection{Dual-iso-B imaging method}
\begin{figure}[h!]
\includegraphics[scale=0.6]{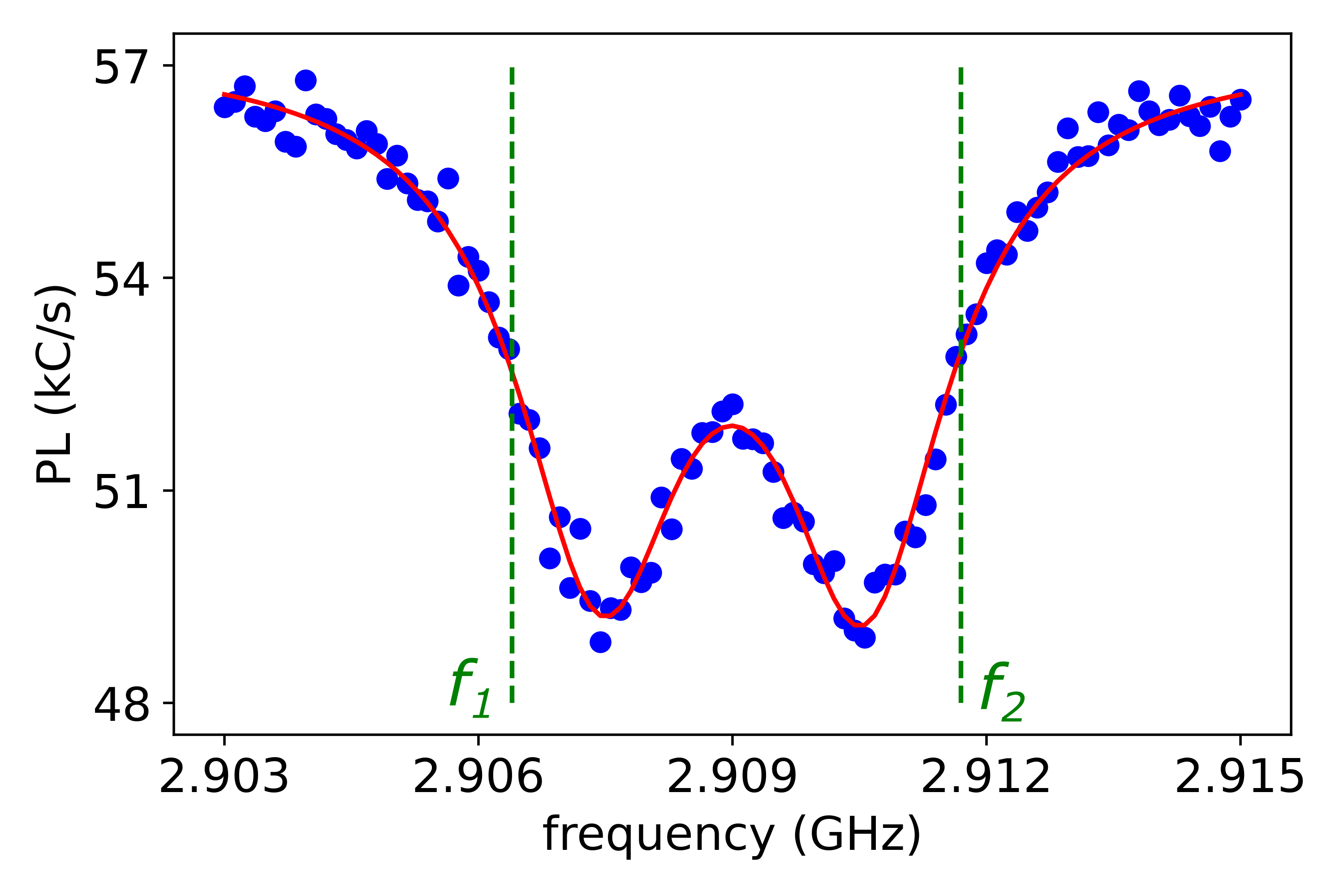}
\caption{\label{fig:singleESR} An ODMR spectrum of the $m_s = 0 \leftrightarrow +1$ electronic spin transition showing two hyperfine resonances corresponding to $m_I = +\frac{1}{2}$ and $-\frac{1}{2}$. $f_1$ and $f_2$ mark the microwave frequencies used during a scan.}
\end{figure}

\begin{figure}[h!]
\includegraphics[scale=0.6]{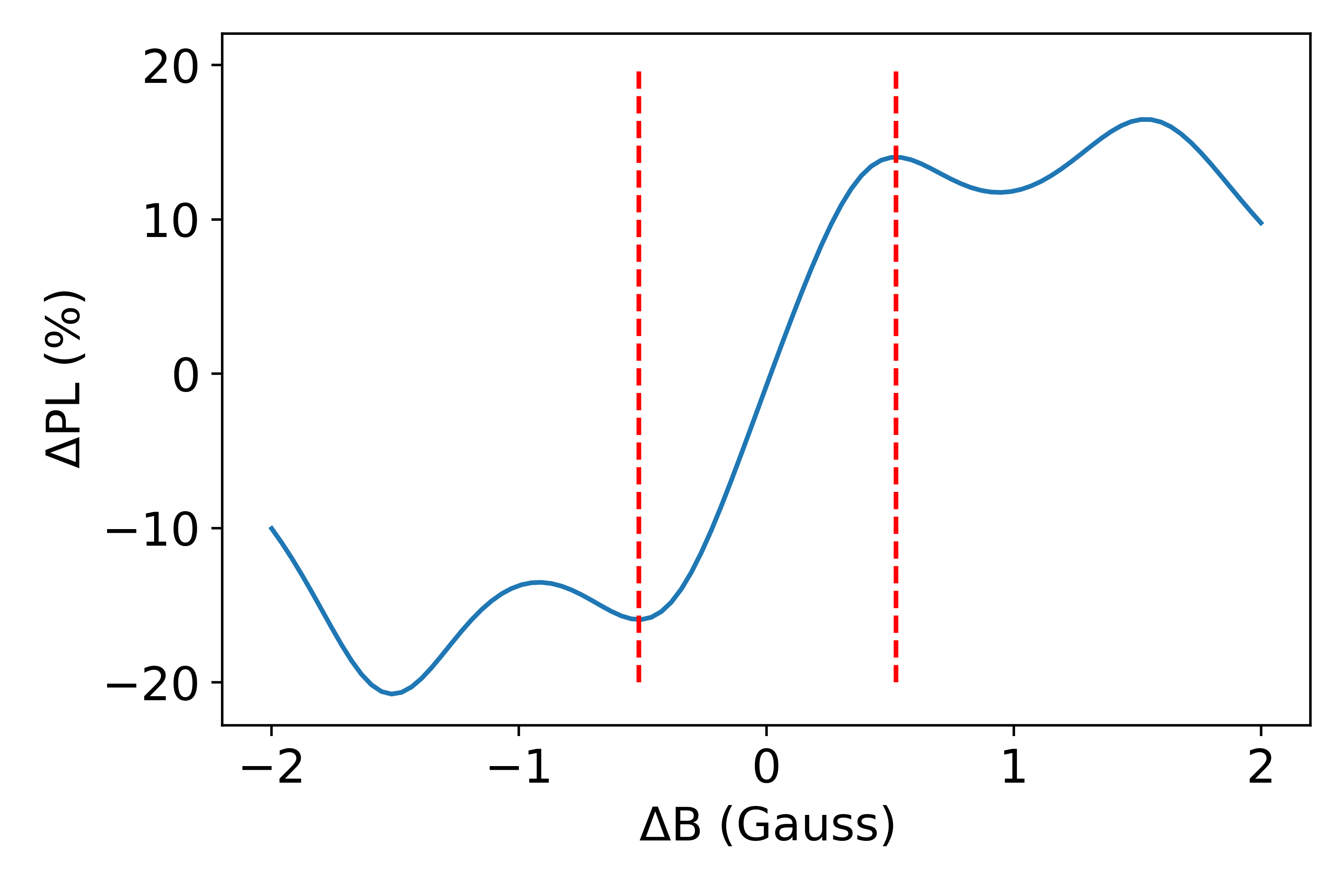}
\caption{\label{fig:dPL} Difference of the PL values at two selected microwave frequencies $f_1$ and $f_2$ as a function of local magnetic field change. The function is approximately linear within ~$\pm$0.5~G. Outside the linear regime, a PL change value may correspond to multiple values of $\Delta B$. }
\end{figure}

For studying current-induced switching (Fig.~3 in the main text), we use the duo-iso-B method\cite{Rondin2012} to determine the local magnetic field. The positions of the dips in the ODMR spectrum linearly shift as a function of field, but the amplitude and linewidth are independent of field for a given NV center. Therefore, we can determine a change in field by finding the displacement of the dips. In the dual-iso-B method, the displacement is determined by sampling the PL at two MW frequencies that are on either side of the ODMR dip. Figure~\ref{fig:singleESR} shows a typical, ODMR spectrum in a bias field of $\sim$20~G.  We select two MW frequencies, $f_1$ and $f_2$, on either side of the resonance at which the corresponding PL values are approximately equal. Figure~\ref{fig:dPL} shows how a difference in PL between $f_1$ and $f_2$ changes as a function of a magnetic field change $\Delta B$. To obtain this prediction, we fit the ODMR dip with the sum of two Lorentzians. We see that the difference in PL is approximately linear for small changes in the field, but becomes multi-valued outside $\sim\pm$0.5~G (red dashed lines in Figure~\ref{fig:dPL}). For the antiferromagnetic sample we study, the magnetic field change is relatively small, and most data points are within the linear regime.  

\subsection{Resonance frequency tracking method}
For studying the dependence of the orientation of the initialization field, we use the resonance frequency tracking method\cite{Schoenfeld2011}. The dual-iso-B method described above is efficient because it samples only at two MW frequencies at each pixel. However, the field range over which a difference in PL can be uniquely associated with a change in field is limited. To increase the dynamic range, we implement a resonance frequency tracking method. At every pixel, we adjust the center of the MW frequencies, $(f_1+f_2)/2$ in steps while keeping the difference $f_2-f_1$ constant until the PL difference is within the shot noise of the measurement. By shifting the center frequency, we always measure within the linear regime unless the magnetic field change between two adjacent pixels exceeds the extent of the linear regime. $\Delta$B at each pixel can be calculated from the final frequencies: $\Delta B = \gamma ((f_1+f_2)/2)-(f_1'+f_2')/2)$, where $f_1$ and $f_2$ are the initial microwave frequencies, $f_1'$ and $f_2'$ are the adjusted final frequencies at each pixel, and $\gamma=2.8$ MHz/G is the gyromagnetic ratio of an NV center. The frequency tracking method allows us to extract the magnetic field quantitatively value at each pixel. 

\subsection{Field initialization procedure}
In the field initialization study, we apply a 1~T magnetic field to the sample using a electromagnet. The sample is mounted between the poles of a electromagnet on a rotation stage with the sample plane parallel to the magnetic field. We slowly ramp the magnetic field from zero to 1 T, and then back to zero. The sample is then moved to the scanning NV center microscope for magnetic imaging. This procedure is repeated 24 times with 15$^\circ$ increments for the full 360$^\circ$ circle. Data are taken in random order to avoid influences from a previous initialization. 

\section{Autocorrelation images}

\begin{figure}[h!]
\includegraphics[scale=0.5]{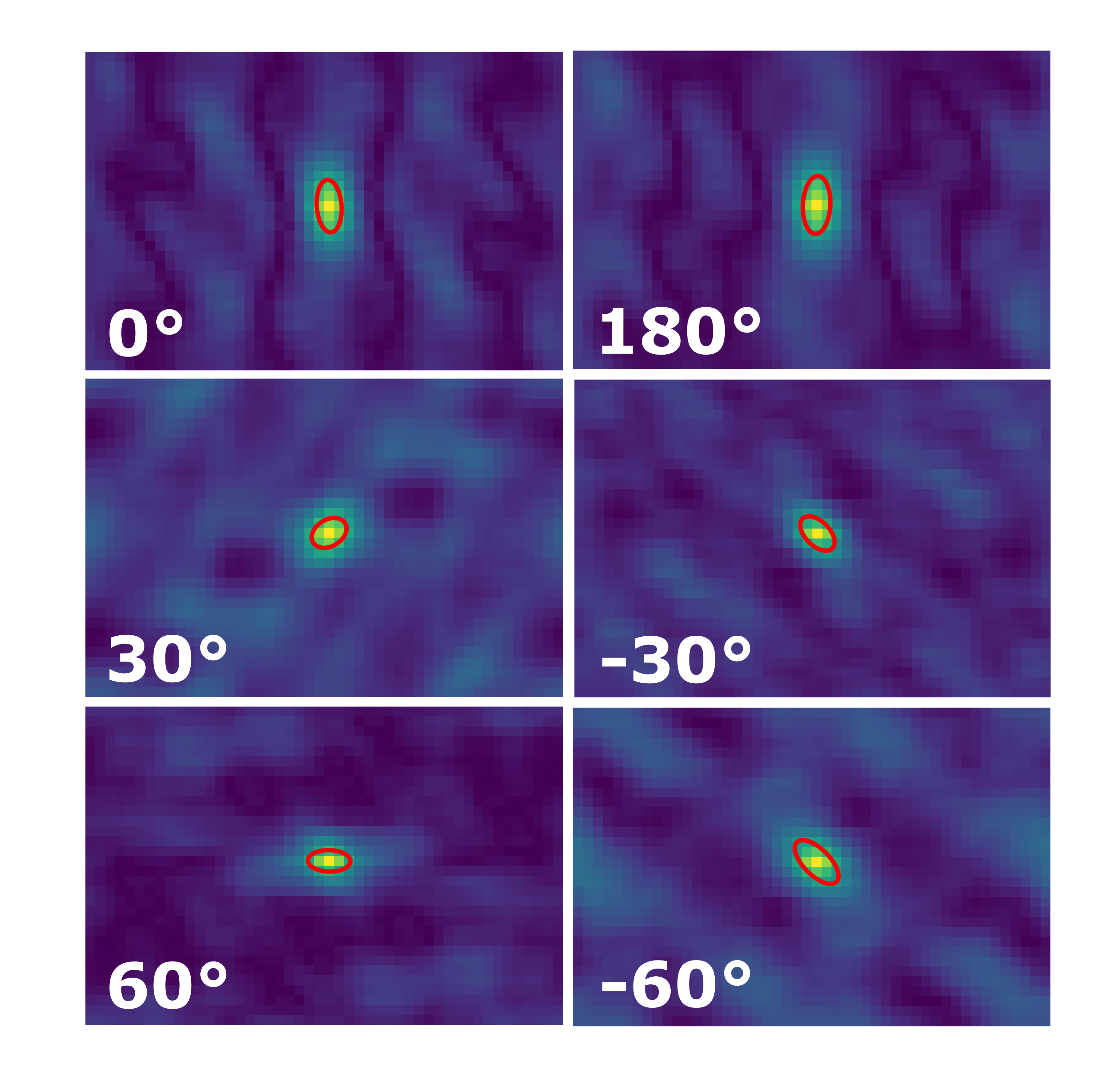}
\caption{\label{fig:SI_ac} Autocorrelation images calculated from magnetic field maps acquired after the sample is initialized with magnetic field oriented at the angles indicated on the images.}
\end{figure}

The two-dimensional autocorrelation of a magnetic image can reveal additional information about magnetic texture \cite{Reith2017,NavaAntonio2021}. The two-dimensional autocorrelation is given by 
\[R (\delta x, \delta y)=\sum_{x,y}I(x, y)\cdot I(x+\delta x, y+\delta y),\] where $R$ is the autocorrelation value, $\delta x$ and $\delta y$ are the displacements from the corresponding x and y, and $I(x, y)$ is the pixel intensity at $(x, y)$. $R$ can be efficiently computed using discrete fast Fourier transform and the Wiener-Khinchin theorem 
\[R(S) = F^{-1}[F(S)\cdot \overline{F(S)}],\]
where $S$ is the 2D map of the signal. Figure~\ref{fig:SI_ac} shows examples of autocorrelation images obtained from the red-boxed regions of the original scans (Fig.3(a) in main text). To characterize the central peak, we set a threshold to define a contour that describes the shape of the central peak. We use a threshold of 30\% of the maximum value in the image. The threshold is chosen so as not to include features outside the central peak, but high enough to preserve the peak's elongation. A linear relationship with R$^2 > 0.95$ can be observed with the threshold ranging from 15\% to 40\% of the maximum. The contour is fitted to an ellipse function
\[r(\theta)=ab/\sqrt{bcos(\theta+\phi)+asin(\theta+\phi)},\]
where $a$ and $b$ are the semi-major and semi-minor axes, and $\phi$ is the ellipse rotation angle. We refer to the rotation angle of the ellipse as the angle of the autocorrelation peak. In Figure 2(c) in the main text we consider a 5-degree constant error for the orientation of the initialization field due to rotation stage accuracy. Standard confidence intervals from the ellipse fitting are used as the y-error-bars. However, additional y-uncertainties are introduced due to the limited number of pixels used for the ellipse fitting. The uncertainty of the ellipse fitting also depends on the choice of threshold and the extent of the fitted area (red boxes). The actual y-error might be larger than the error bar shown. 

\section{Additional data}
\subsection{Field initialization on device B}
\begin{figure}[h!]
\includegraphics[scale=0.6]{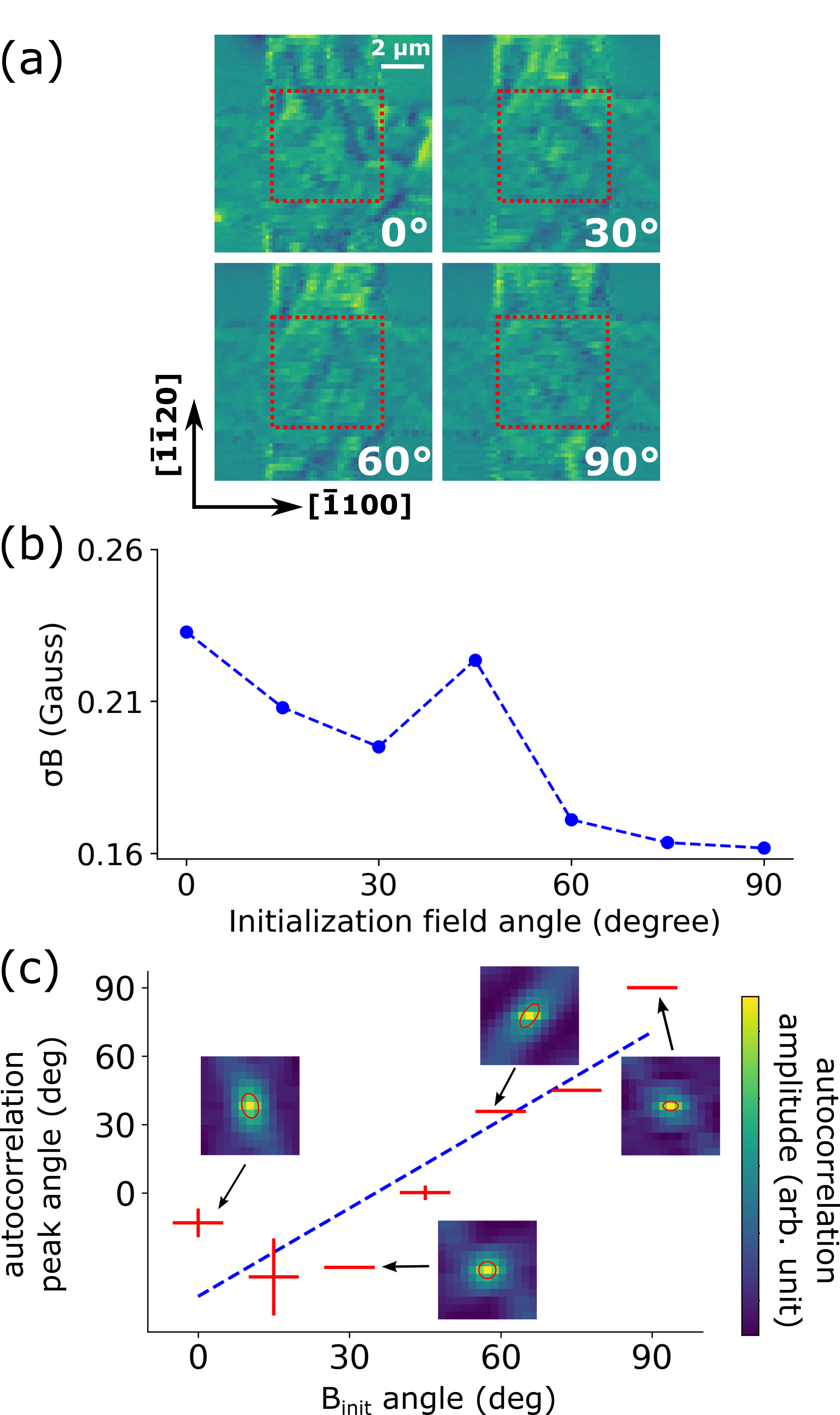}
\caption{\label{fig:SI_B4} (a) Magnetic field images acquired on device B after initialization with a magnetic field at the angles indicated on the images. Autocorrelation images are calculated from the red-boxed regions. (b) Magnetic field standard deviations of the post-initialization images. The standard deviation decreases from 0$^\circ$ to 90$^\circ$, which agrees with the trend on device A. (c) The fitted autocorrelation peak angle as a function of the initialization field angle. The linear fit has a slope of 1.29 and an offset of -45.5$^\circ$, with an R$^2$ value of 0.81.}
\end{figure}

We perform field initialization for the first 90$^\circ$ on device B. The magnetic field maps and their corresponding autocorrelation images are shown in Figure \ref{fig:SI_B4}. The results show that device B behaves similar to device A: the standard deviation of the magnetic field shows an overall decreasing trend from 0$^\circ$ to 90$^\circ$ (Fig.~\ref{fig:SI_B4}(b)), and the autocorrelation peak angle has a linear relationship with the initialization field angle (Fig.~\ref{fig:SI_B4}(c)). The linear fit has R$^2>0.75$ for a threshold ranging from 30\% to 40\% of the maximum. A threshold of 35\% is used for Figure~\ref{fig:SI_B4}. The weaker linear correlation and smaller threshold range on device B compared to device A can be explained by the difference in the device size. Since the size of device B is much smaller than device A, its resulting magnetic states are likely more affected by the edge of the device. We use a higher threshold to exclude effects from the device edge as much as possible, however, the higher threshold leaves fewer pixels available for the ellipse fitting. The small number of pixels used in the ellipse fitting may introduce additional uncertainty that is not captured by the standard error analysis that we use to generate the error bars. 

\subsection{Effects from the NV axis orientation}
\begin{figure}[h!]
\includegraphics[scale=0.8]{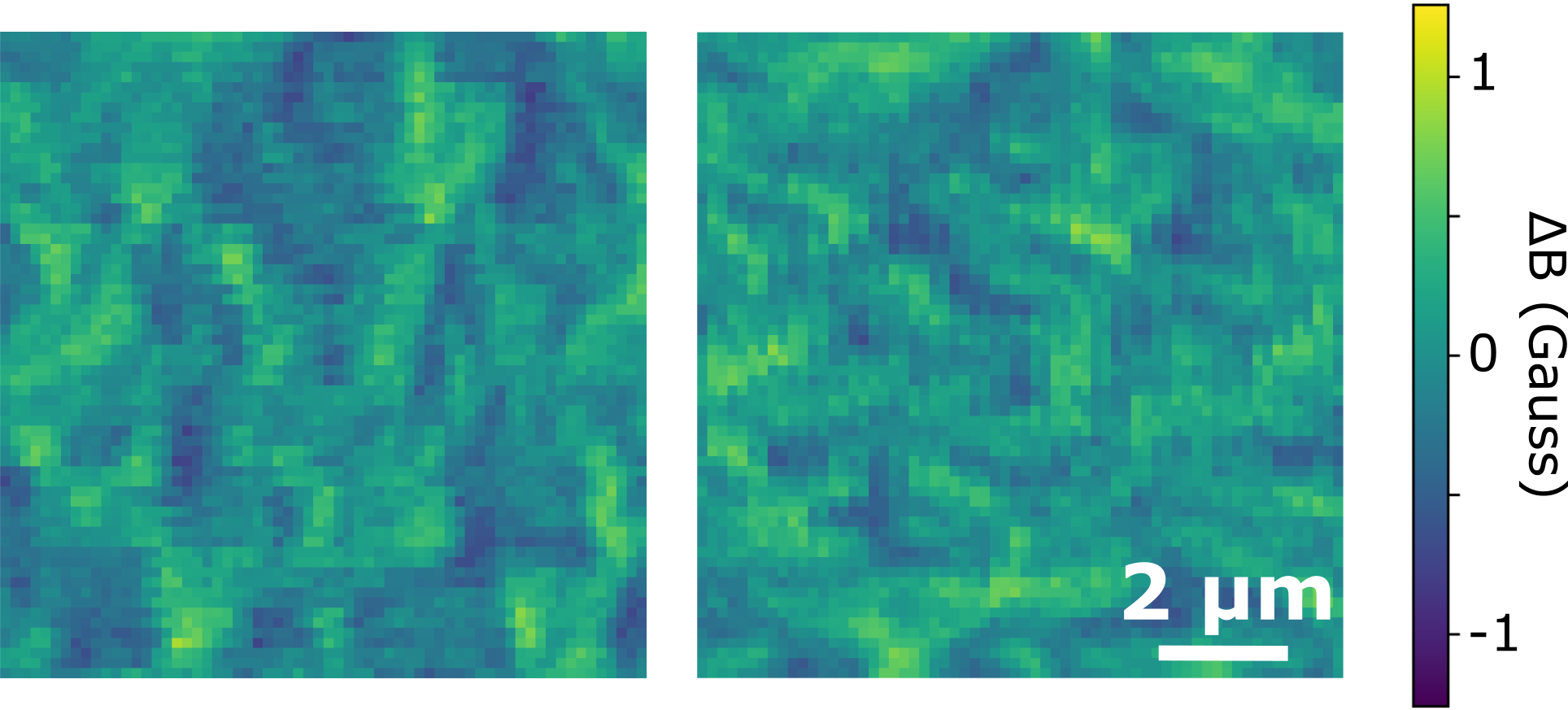}
\caption{\label{fig:SI_90rot} Magnetic images of the same area on device A, with an NV axis in-plane projection along $[\bar{1}100]$ (left) or $[\bar{1}\bar{1}20]$ (right). }
\end{figure}

The diamond NV center probe used in this experiment has an N-V axis oriented at a 54$^{\circ}$ angle with respect to the sample plane normal and an in-plane projection along $[\bar{1}100]$. To confirm that the peaks of the magnetic field standard deviation (Fig.~\ref{fig:SI_B4}(B) and Fig.~2(B) in main text) are not influenced by the N-V axis orientation, we scan the same region of the sample with the sample rotated by 90 degrees in-plane (i.e.~N-V axis in-plane projection along $[\bar{1}\bar{1}20]$). Figure~\ref{fig:SI_90rot} shows two magnetic images taken on the same region of device A. The sample is initialized with an 0$^\circ$ initialization field. The two images show similar magnetic structure and have a similar magnetic field standard deviation of 0.28 (unrotated) and 0.26 (rotated).

\subsection{25 mA current pulse switching}
\begin{figure}[h!]
\includegraphics[scale=0.6]{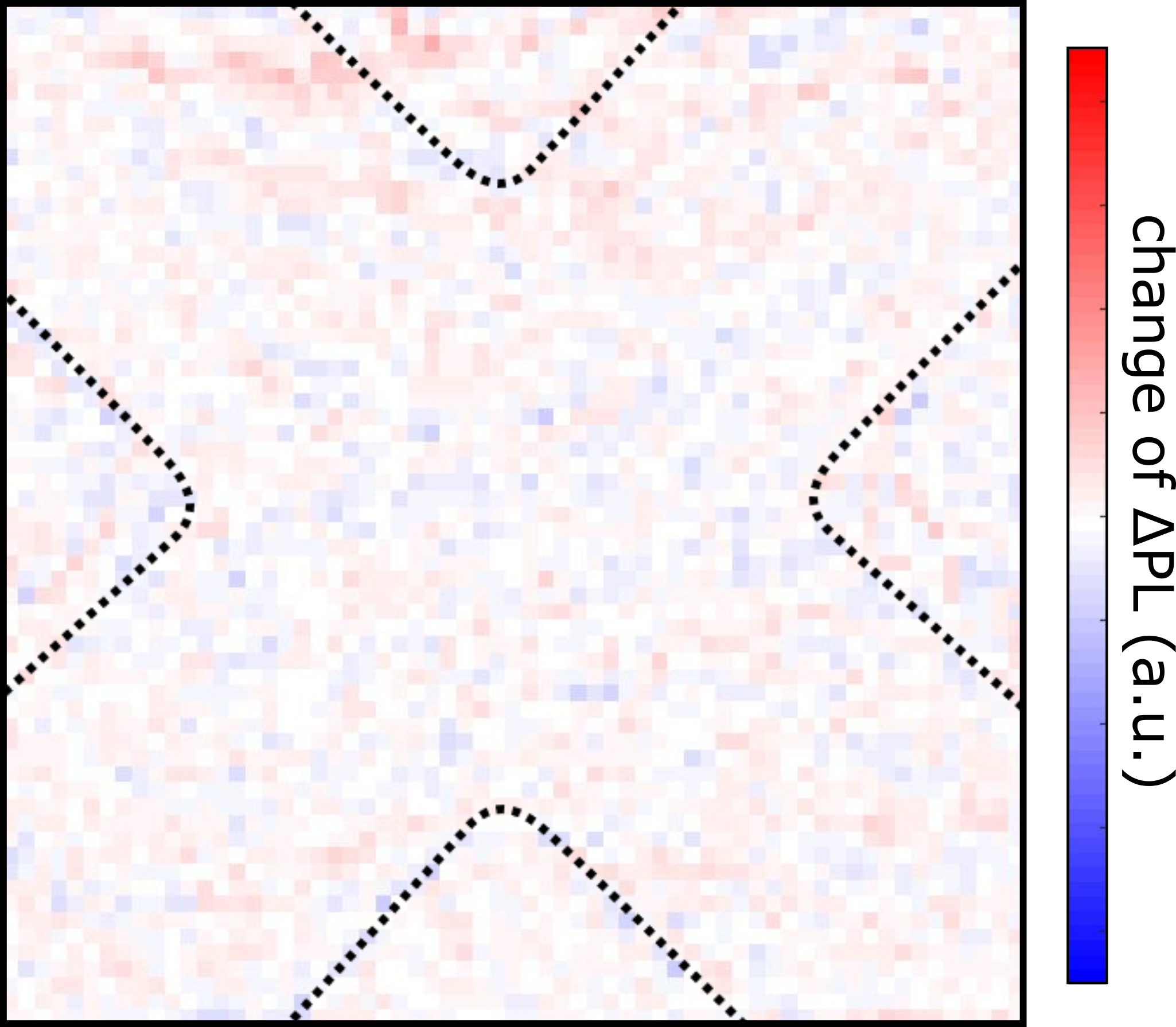}
\caption{\label{fig:25mAscan} Difference image after passing a \SI{1}{\us}, 25 mA current pulse through the device. No switching is observed.}
\end{figure}

On the additional device, we pass a current pulse of a \SI{1}{\us}, 25 mA current pulse ($j = 8.3 \times 10^{11} A/m^2$), which is the maximum current density we can achieve with our instrumentation. The difference image shows no sign of current-induced switching (Fig.~\ref{fig:25mAscan}). 

\subsection{Switching outside current path}
\begin{figure}[h!]
\includegraphics[scale=0.6]{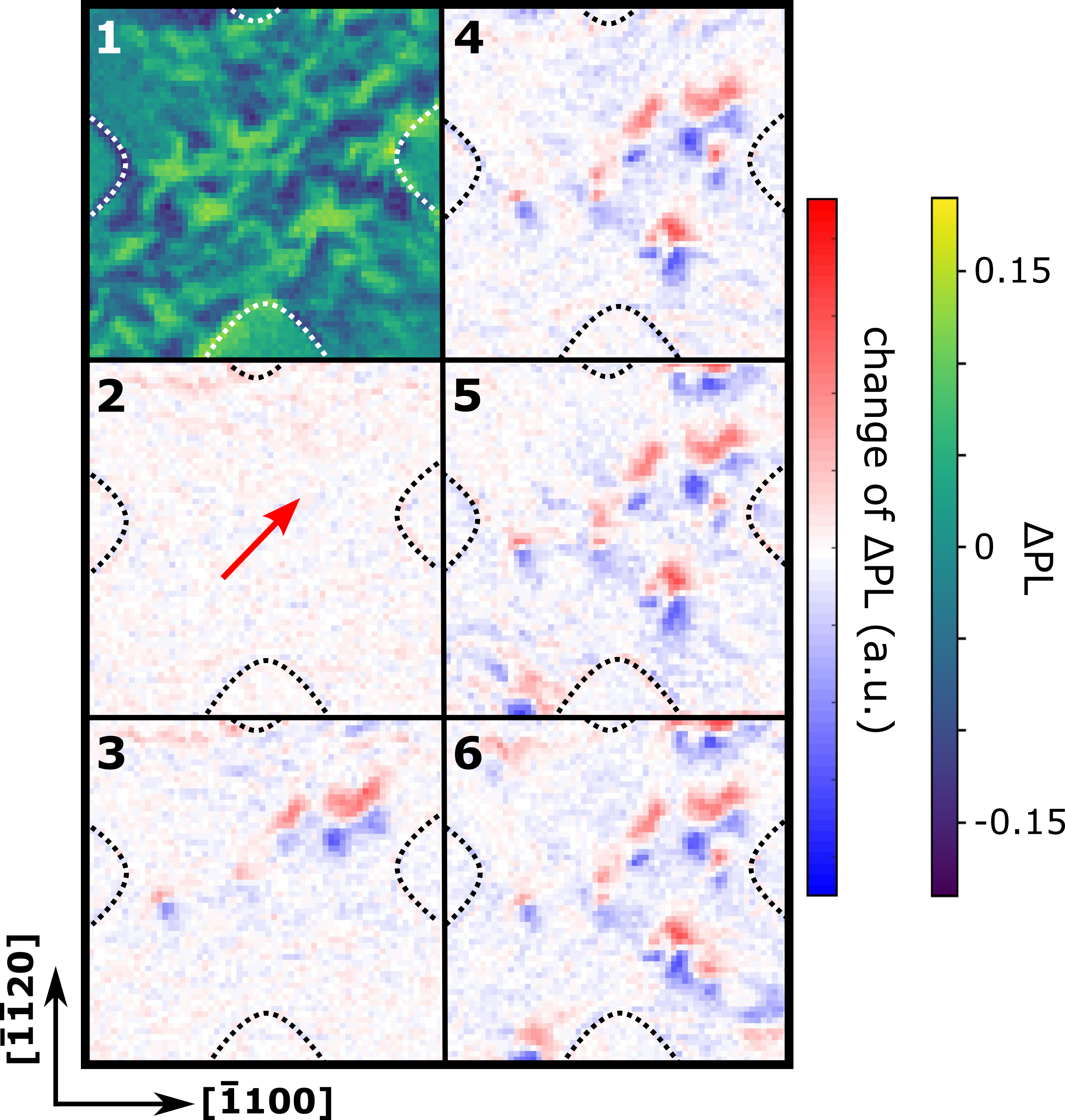} 
\caption{\label{fig:SI_B3} Magnetic field images acquired using the dual-iso-B method on additional device. (1) Image before applying any current pulse. (2-6) Difference images obtained by subtracting the scan in (1) from the NV scan taken after passing \SI{100}{\us} current pulses with amplitude of (2) 3~mA, (3) 7~mA, (4) 13~mA, (5) 16~mA, and (6)~21 mA. The red arrow indicates the current pulse direction.}
\end{figure}

We observe current-induced switching on an additional device on the same sample as device B. This set of measurements is made without initializing the device in a magnetic field. The orientation of this device is rotated by 45$^\circ$ compared to the cross-shaped device B in the main text. Current pulses are passed from the bottom left leg to the top right leg. We find similar switching behavior in this device, and switching outside the current path can be observed on its bottom right leg (Fig.~\ref{fig:SI_B3}). 

\bibliography{SI_ref}